\documentclass[useAMS,usenatbib]{mn2e}
\pdfpagewidth=\paperwidth
\pdfpageheight=\paperheight

\usepackage{graphicx,mathtools}
\usepackage{amsmath,amssymb}
\usepackage[dvipsnames]{xcolor}
\usepackage{xparse,xcolor}
\usepackage{xspace}
\usepackage{natbib}
\usepackage{fixltx2e}

\makeatletter
\def\Ginclude@eps#1{%
 \message{<#1>}%
  \bgroup
  \def\@tempa{!}%
  \dimen@\Gin@req@width
  \dimen@ii.1bp%
  \divide\dimen@\dimen@ii
  \@tempdima\Gin@req@height
  \divide\@tempdima\dimen@ii
    \includegraphics{#1}%
  \egroup}
\makeatother

\font\gkvec=cmmib10                        %for boldface lowercase Greek
\def\balpha{\hbox{{\gkvec\char11}}}        %bold face alpha
\def\bbeta{\hbox{{\gkvec\char12}}}         %bold face beta
\def\bgamma{\hbox{{\gkvec\char13}}}        %bold face gamma
\def\bdelta{\hbox{{\gkvec\char14}}}        %bold face delta
\def\bepsilon{\hbox{{\gkvec\char15}}}      %bold face epsilon
\def\bzeta{\hbox{{\gkvec\char16}}}         %bold face zeta
\def\boldeta{\hbox{{\gkvec\char17}}}       %bold face eta
\def\btheta{\hbox{{\gkvec\char18}}}        %bold face theta
\def\biota{\hbox{{\gkvec\char19}}}         %bold face iota
\def\bkappa{\hbox{{\gkvec\char20}}}        %bold face kappa
\def\blambda{\hbox{{\gkvec\char21}}}       %bold face lambda
\def\bmu{\hbox{{\gkvec\char22}}}           %bold face mu
\def\bnu{\hbox{{\gkvec\char23}}}           %bold face nu
\def\bxi{\hbox{{\gkvec\char24}}}           %bold face xi
\def\bpi{\hbox{{\gkvec\char25}}}           %bold face pi
\def\brho{\hbox{{\gkvec\char26}}}          %bold face rho
\def\bsigma{\hbox{{\gkvec\char27}}}        %bold face sigma
\def\btau{\hbox{{\gkvec\char28}}}          %bold face tau
\def\bupsilon{\hbox{{\gkvec\char29}}}      %bold face upsilon
\def\bphi{\hbox{{\gkvec\char30}}}          %bold face phi
\def\bchi{\hbox{{\gkvec\char31}}}          %bold face chi
\def\bpsi{\hbox{{\gkvec\char32}}}          %bold face psi
\def\bomega{\hbox{{\gkvec\char33}}}        %bold face omega

\newcommand{\have}[1]{\left\langle #1 \right\rangle}

\newcommand{\bcave}[1]{\left\lbrace #1 \right\rbrace_t}

\title[Convective Quenching of Field Reversals]{Convective Quenching of Field Reversals in Accretion Disc Dynamos}
\author[M. S. B. Coleman et al. ]{
\parbox{\textwidth}{
Matthew S. B. Coleman$^{1}$\thanks{E-mail: mcoleman@physics.ucsb.edu},
Evan Yerger$^{1,2}$,
Omer Blaes$^{1,3}$,
Greg Salvesen$^{1}$\thanks{NSF Astronomy \& Astrophysics Postdoctoral Fellow.},
and Shigenobu Hirose$^{4}$\vspace{0.4cm}
}\\
\parbox{\textwidth}{
$^{1}$ Department of Physics, University of California, Santa Barbara, CA
93106, USA\\
$^{2}$ Department of Astrophysical Sciences, Program in Plasma Physics, Princeton, NJ 08540, USA\\
$^{3}$ Kavli Institute for Theoretical Physics, University of California,
Santa Barbara, CA 93106, USA\\
$^{4}$ Department of Mathematical Science and Advanced Technology, Japan
Agency for Marine-Earth Science and Technology, Yokohama, Kanagawa 236-0001, Japan
}
}

\begin{document}

\date{Accepted ---. Received ---; in original form ---}

\pagerange{\pageref{firstpage}--\pageref{lastpage}} \pubyear{2016}

\maketitle

\label{firstpage}

\begin{abstract}
Vertically stratified shearing box simulations of magnetorotational turbulence
commonly exhibit a so-called butterfly diagram of quasi-periodic azimuthal
field reversals.  However, in the presence of hydrodynamic convection, field
reversals no longer occur.  Instead, the azimuthal field strength fluctuates
quasi-periodically while maintaining the same polarity, which can
either be symmetric or antisymmetric about the disc midplane.
Using data from the simulations
of \citet{HIR14}, we demonstrate that the lack of field reversals in the presence
of convection is due to hydrodynamic mixing of magnetic field from the
more strongly magnetized upper layers into the midplane, which then annihilate
field reversals that are starting there.
Our convective simulations differ in several respects from those reported in previous work by others, in which stronger magnetization likely plays a more important role than convection.
\end{abstract}

\begin{keywords}
accretion, accretion discs, dynamos, convection --- MHD --- turbulence --- stars: dwarf novae.
\end{keywords}

\section{Introduction}

As a cloud of gas contracts under the influence of gravity, it is likely to reach a point where net rotation dominates the dynamics and becomes a bottleneck restricting further collapse. This scenario naturally leads to a disc structure, thus explaining why accretion discs are so prevalent in astrophysics. The question of how these discs transport angular momentum to facilitate accretion still remains. For sufficiently
electrically conducting
discs, there is a reasonable consensus that the magnetorotational instability (MRI) is the predominate means of transporting angular momentum, at least on local scales \citep{BAL98}. There has also been significant work on understanding
non-local mechanisms of angular momentum transport, such as spiral waves \citep{JU16}.  While these global structures may be important in discs with low conductivity, local MRI simulations of fully ionized accretion discs produce values of $\alpha$ consistent with those inferred from
observations for accretion discs in binary systems.
This is even true for the case of dwarf novae, for which MRI simulations
lacking net vertical magnetic flux previously had trouble with matching
observations \citep{KIN07,K12}.
This is because hydrodynamic convection
occurs in the vicinity of the hydrogen ionization regime, and enhances
the time-averaged \citet{SS73} alpha-parameter \citep{HIR14,HIR15}.
Enhancement of MRI turbulent stresses by convection was also independently claimed by \citet{BOD12,BOD15}.
Whether this enhancement is enough to reproduce observed dwarf nova light curves remains
an unanswered question, largely due to uncertain physics in the quiescent
state and in the propagation of heating and cooling fronts \citep{COL16}.

The precise reason as to why convection enhances the turbulent stresses
responsible for angular momentum transport is still not fully understood.
\citet{BOD12,BOD13} conducted simulations with fixed thermal diffusivity and impenetrable vertical boundary conditions, and
found that when this diffusivity was low, the time and horizontally-averaged
vertical density profiles became very flat or even slightly inverted, possibly due to
hydrodynamic convection taking place.  They
suggested that either these flat profiles, or the overturning convective
motions themselves, might make the magnetohydrodynamic dynamo more efficient.
\citet{HIR14} used radiation MHD simulations with outflow vertical boundary conditions, and the hydrogen ionization
opacities and equation of state that are relevant to dwarf novae. They found
intermittent episodes of convection separated by periods of radiative
diffusion.  The beginning of the convective episodes were associated with
an enhancement of energy in vertical magnetic field relative to the horizontal
magnetic energy, and this was then followed by a rapid growth of
horizontal magnetic energy.  These authors therefore suggested that
convection seeds the axisymmetric magnetorotational instability, albeit in
a medium that is already turbulent.  In addition, the phase lag between
stress build up and heating which causes pressure to build up also contributes
to an enhancement of the alpha parameter.
The mere presence of vertical hydrodynamic convection is not sufficient
to enhance the alpha parameter, however;  the Mach number of the convective
motions also has to be sufficiently high, and in fact the alpha parameter
appears to be better correlated with the Mach number of the convective motions
than with the fraction of heat transport that is carried by convection
\citep{HIR15}.

Hydrodynamic convection does not simply enhance the turbulent stress
to pressure ratio, however.
It also fundamentally alters the character of the MRI dynamo.  In the
standard weak-field MRI, vertically stratified shearing box simulations
exhibit
quasi-periodic field reversals of the azimuthal magnetic field
($B_y$) with periods of $\sim 10$ orbits \citep{BRA95,DAV10}. These reversals start near the
midplane and propagate outward making a pattern (see top-left panel of
Figure \ref{fig:By} below) which resembles a time inverse of the solar sunspot
butterfly diagram.
The means by which these field reversals propagate away from the midplane is
likely the buoyant advection of magnetic flux tubes
\citep{GRE10}, and many studies have also suggested that magnetic buoyancy is important in accretion discs \citep[e.g.][]{GAL79,BRA98,MIL00,HIR06,DAV10,BLA11}. Magnetic buoyancy
is consistent with the Poynting flux which tends to be oriented
outwards (see top-right panel of Fig. \ref{fig:By}), and we
give further evidence supporting this theory below.
While this explains how field reversals propagate through the disc, it does not explain how these magnetic field reversals occur in the first place, and despite numerous dynamo models there is currently no consensus on the physical mechanism driving the reversals \citep[e.g.][]{BRA95,GRE10,SHI10,SHI16,S&B_2015}.

However, in the presence of convection,
the standard pattern of azimuthal field reversals is disrupted.
Periods of convection appear to be characterized by longer term
maintenance of a particular azimuthal field polarity, and this persistent
polarity can be of even \citep{BOD15} or odd parity with respect to the disk midplane.
As we discuss in this paper, the simulations of \citet{HIR14} also exhibit
this pattern of persistent
magnetic polarity during the intermittent periods of convection, but the
field reversals associated with the standard butterfly diagram return during
the episodes of radiative diffusion (see Fig.~\ref{fig:By} below).
Here we exploit this intermittency to
try and understand the cause of the persistent magnetic polarity in the
convective episodes.  We demonstrate that this is due to hydrodynamic
mixing of magnetic field from strongly magnetized regions at high altitude
back toward the midplane.

This paper is organized as follows.
In Section 2 we discuss the butterfly diagram in detail and how it changes character when convection occurs. In Section 3 we describe magnetic buoyancy and the role it plays in establishing the butterfly diagram,
and the related thermodynamics.
We explain how convection acts to alter these effects in Section 4. The implications of this work are discussed in Section 5, and our results are summarized in Section 6.

\begin{figure*}
	\includegraphics[width=.495\linewidth]{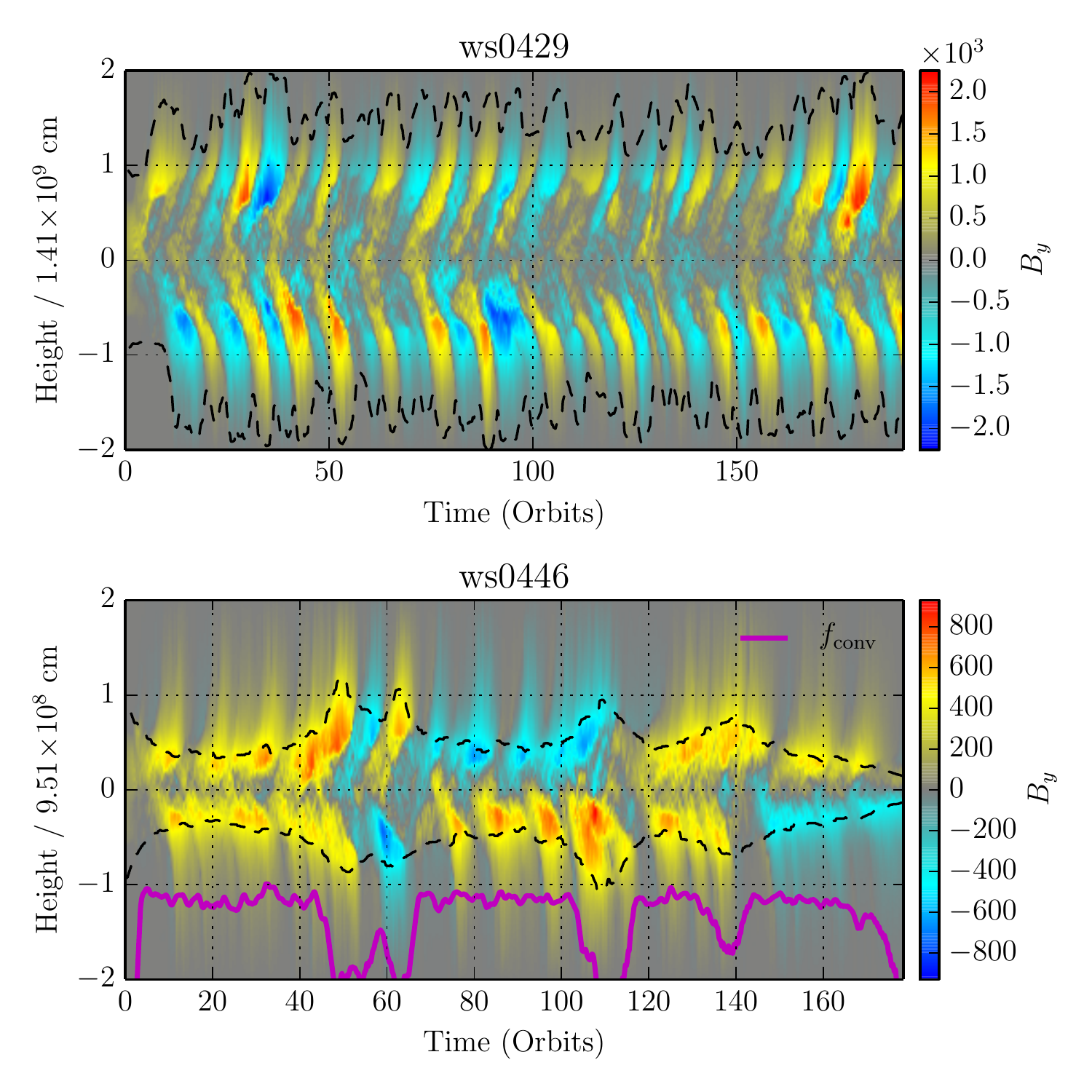} \hfill
	\includegraphics[width=.495\linewidth]{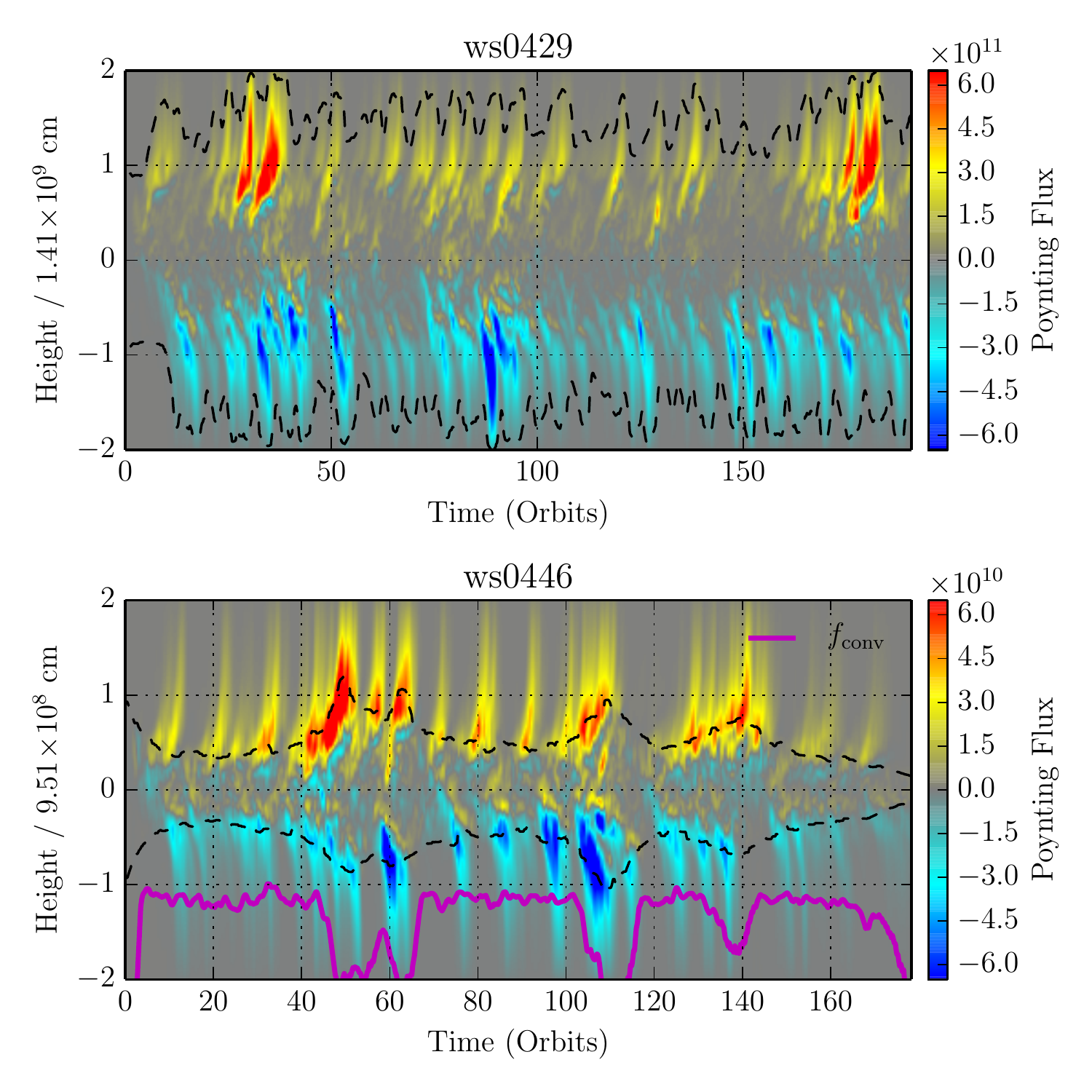}
	\caption{Horizontally-averaged azimuthal magnetic field $\have{B_y}$ (left frames), and 1 orbit smoothed horizontally-averaged vertical Poynting flux $\bcave{\have{F_{{\rm Poynt},z}}}$ (right frames) as a function
		of time and height for a radiative simulation (ws0429, top frames) and a simulation which exhibits convective epochs (ws0446, bottom frames). For the Poynting flux frames, we have purposely allowed some of the data to lie outside the colorbar range (which then shows up as saturated regions) in order to increase the color contrast in the
midplane regions. The normalizations indicated in the vertical axes are the respective
simulation length units.
		The dashed black lines show the time-dependent heights of
		the photospheres in the horizontally averaged structures.
		For simulation ws0446, which exhibits convection, the convective fraction $f_{\rm conv}-2$ (see Eqn. \ref{eqn:fconv}) is plotted in magenta. Note that $f_{\rm conv}-2$ uses the same vertical scale as $B_y$, i.e. when the magenta line is near $-1$ then $f_{\rm conv}\approx 1$.
		Focusing on $B_y$, the radiative simulation ws0429 shows the standard pattern of field reversals normally associated with the butterfly diagram. In simulation ws0446 where $f_{\rm conv}$ is high, the field maintains its sign and all changes in sign/parity are associated with a dip in $f_{\rm conv}$.
	}
	\label{fig:By}
\end{figure*}

\section{The Butterfly Diagram}

\begin{table*}
\begin{tabular}{lrrrrrrrrrrrrr}
\hline
\hline \\[-2.5ex]
Simulation & $h_0$ & $\Sigma$ & $T_\text{eff}$ &
$\alpha$ &  $N_x$ & $N_y$ & $N_z$ &$L_x/h_0$ & $L_y/h_0$ & $L_z/h_0$ &
$L_z/h_\text{p}$ & $t_\text{th}$\\
\hline
\hline \\[-2.5ex]
ws0446 (conv) & 9.51E+08 & 127 & 7490 & 0.1062 & 32 & 64 & 256 & 0.500 & 2.000 & 4.000 & 12.0 & 5.06 \\
ws0429 (rad) & 1.41E+09 & 1030 & 13352 & 0.0332 & 32 & 64 & 256 & 0.500 & 2.000 & 4.000 & 8.54 & 9.49 \\
\end{tabular}
\caption{Simulation parameters for the convective simulation ws0446 and radiative simulation ws0429. The units of time averaged surface
	densities ($\Sigma$), effective temperatures
	($T_\text{eff}$), height ($h_0$), and thermal time ($t_\text{th}$) are, respectively, g cm$^{-2}$, K, cm, and orbits. $L_x$, $L_y$,
	and $L_z$ are the lengths, and $N_x$, $N_y$, and $N_z$ are the numbers of cells,
	in the $x$, $y$, and in $z$ directions, respectively.
	The pressure scale height of the steady state is computed as $h_\text{p} \equiv \int\left[\left<p_\text{thermal}\right>\right]dz / 2\max(\left[\left<p_\text{thermal}\right>\right]) $. }
\label{tab:param}
\end{table*}

To construct the butterfly diagram and explore its physical origin, it is useful to
define the following quantities related to some fluid variable $f$:  the horizontal
average of this quantity, the variation with respect to this horizontal average,
and a version of the variable that is smoothed in time over one orbit.  These are
defined respectively by
\begin{subequations}
\begin{align}
\have{f}(t,z) &\equiv\frac{1}{L_xL_y}\int_{-L_x/2}^{L_x/2}dx\int_{-L_y/2}^{L_y/2}dy
f(t,x,y,z)\\
\label{eqn:delta-f}
\delta\! f &\equiv f - \have{f},\\
\bcave{f}(t) &\equiv\left.\int_{t-1/2}^{t+1/2} f(t^\prime) \, dt^\prime \middle/ 1\text{ orbit}\right..
\end{align}
\end{subequations}
Here $L_x$, $L_y$, and $L_z$ are the radial, azimuthal and vertical extents of the simulation domain, respectively (listed in Table \ref{tab:param}). Additionally we define the quantity $f_{\rm conv}$ as a means to estimate the fraction of vertical energy transport which is done by convection:
\begin{equation}
\label{eqn:fconv}
f_{\rm conv}(t) \equiv \left\{\dfrac{\int\left\{\left<\left(e+E\right)v_z\right>\right\}_t{\rm sign}(z)\left<P_\text{th}\right>dz}{\int\left\{\left<F_{{\rm tot},z}\right>\right\}_t{\rm sign}(z)\left<P_\text{th}\right>dz}\right\}_t,
\end{equation}
where $e$ is the gas internal energy density, $E$ is the radiation energy density, $v_z$ is the vertical velocity, $P_{\rm th}$ is the thermal pressure (gas plus radiation), and $F_{{\rm tot},z}$ is the total energy flux in the vertical direction, including Poynting
and radiation diffusion flux.
These quantities will assist us in analyzing and discussing the interactions between convection and dynamos in accretion discs.

\begin{figure}
	\includegraphics[width=1.0\linewidth]{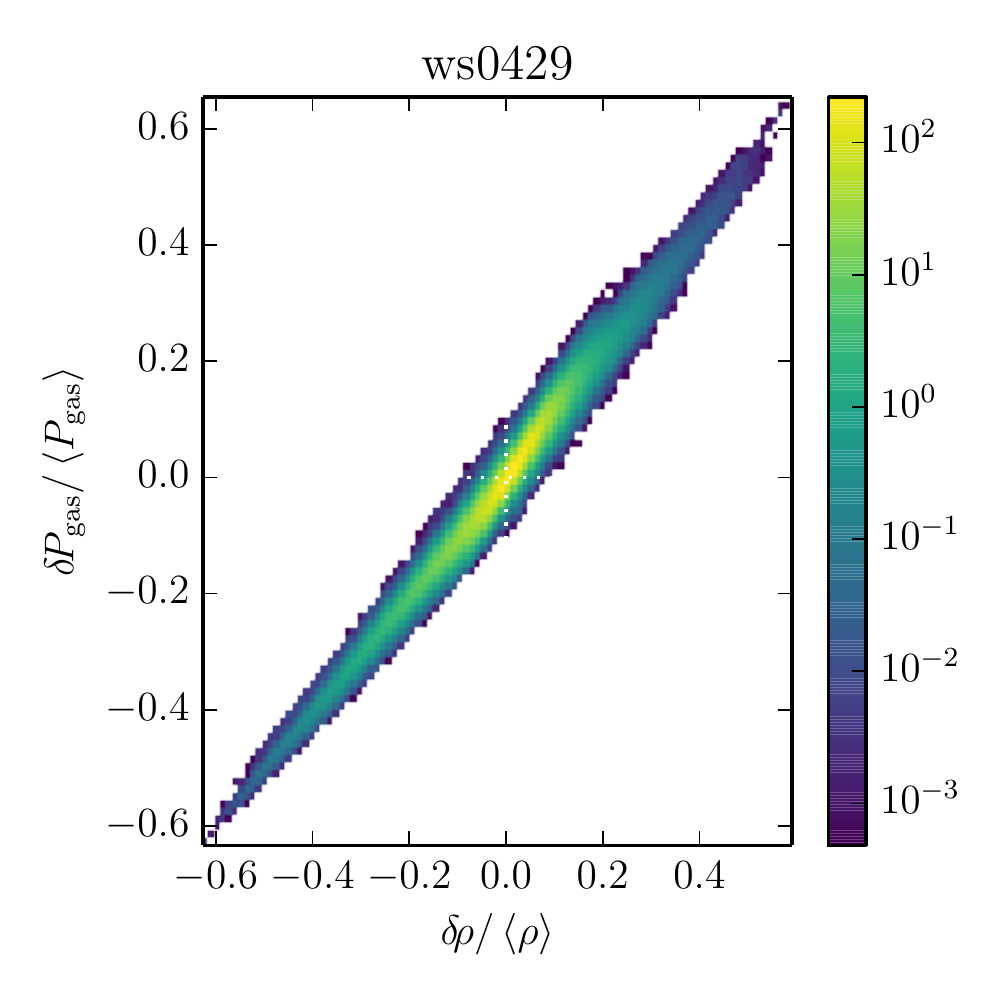}
	\caption{
		A 2D histogram of horizontal variations (see Eqn. \ref{eqn:delta-f}) of mass density ($\delta\rho$, horizontal axis) and gas pressure ($\delta P_{\rm gas}$, vertical axis) for the fully radiative simulation ws0429.
		Each grid zone saved to file from this simulation with $90\leq t \leq190$ and $|z|\leq 0.5$ simulation units is placed into one of $100\times 100$ bins based on its local properties.
		The resulting normalized probability density is shown here with the given color bar.
		The probability density, $p$, is normalized so that $\int p(\tilde{x},\tilde{y})\, d\tilde{x}\,d\tilde{y}=1$ where $\tilde{x}$ and $\tilde{y}$ are the variables used for the horizontal and vertical axes, respectively.
		The tight linear correlation between $\delta\rho$ and $\delta P_{\rm gas}$ signifies that temperature variations are small at any fixed height,
i.e. horizontal fluctuations are isothermal.
		}
	\label{fig:pgas-rho-rad}
\end{figure}

\begin{figure*}
	\includegraphics[width=1.0\linewidth]{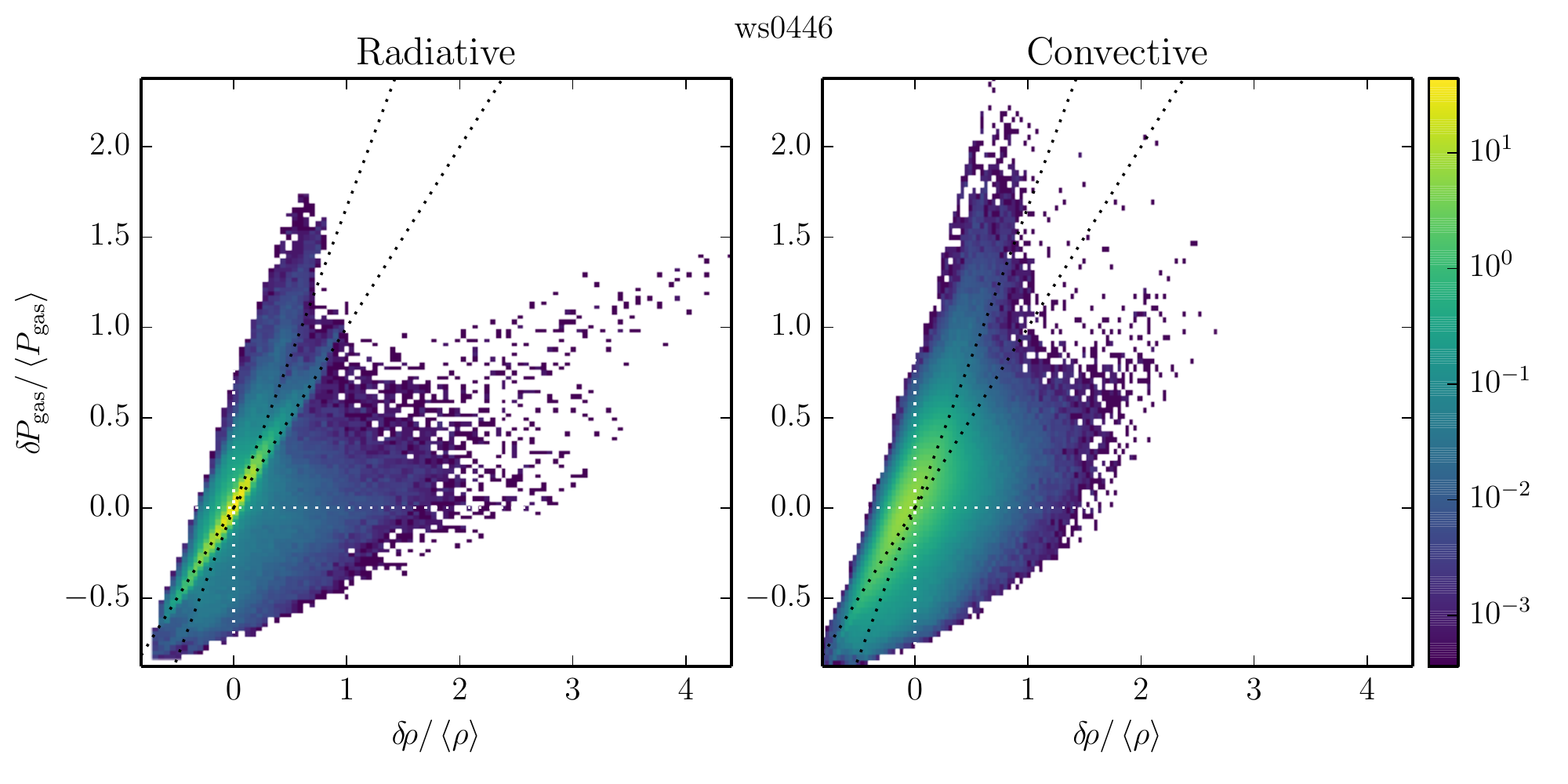}
	\caption{
		2D histograms of horizontal variations of mass density ($\delta\rho$,
horizontal axis) and gas pressure ($\delta P_{\rm gas}$, vertical axis) in the
convective simulation ws0446.
		The left panel shows data from a radiative epoch ($49\leq t \leq 64$),
and the right panel shows data from a convective epoch ($70\leq t \leq 100$). The two dotted black lines in each figure show the expected relations for
isothermal (slope 1) and monatomic adiabatic (slope 5/3) gases.
		During the radiative epoch there is a significant population of cells that lie on the line $\delta P_{\rm gas}=\delta\rho$ just as in Fig. \ref{fig:pgas-rho-rad}. However there is also a significant amount of dispersion which resembles the probability distribution of the convective epoch, therefore it is likely that this dispersion is a remnant of the previous convective epoch. During the convective epoch the yellow core of the probability distribution follows a linear trend with a slope steeper than 1. This is indicative of perturbations which are adiabatic and the spread in slopes over this yellow region are likely due to the variation in $\Gamma_1\equiv(\partial\ln P_{\rm gas}/\partial\ln\rho)_s$ from our equation of state, which includes in particular the effects of ionizing hydrogen.
Values of $\Gamma_1$ range from about 1.15 to 5/3 in this simulation. The population below
$\delta P_{\rm gas}=\delta\rho$ is likely a result of vertical mixing.
	}
	\label{fig:pgas-rho-conv}
\end{figure*}

The butterfly diagram is obtained by plotting $\have{B_y}$ as a function of time and distance from the disc midplane (see left frames of Fig. \ref{fig:By}). The radiative simulation ws0429 \citep[from][and listed in Table \ref{tab:param}]{HIR14} shows the standard pattern of field reversals normally associated with the butterfly diagram, which appear to start at the midplane and propagate outwards. This outward propagation of magnetic field is consistent with the Poynting flux (also shown in Fig. \ref{fig:By}), which generally points outwards away from the midplane.

When simulations with convection are examined (e.g. ws0446 listed in Table \ref{tab:param}), however, the butterfly diagram looks
completely different (see bottom left frame of Fig. \ref{fig:By}), as first discussed
by \citet{BOD15}. Similar to the lack of the azimuthal magnetic field reversals found
by these authors, we find that when convection is present in the \citet{HIR14}
simulations, there is also a lack of field reversals.
Additionally, we find that the azimuthal magnetic field in the
high altitude ``wings"  of the butterfly diagram
is better characterized by quasi-periodic pulsations, rather
than quasi-periodic field reversals.  These pulsations have roughly the same period as the field reversals found
in radiative epochs.
For example, the convective simulation ws0446 shown in Fig.~\ref{fig:By} has a radiative
epoch where field reversals occur (centered near 55 orbits), and the behavior of this
epoch resembles that of the radiative simulation ws0429. However, during convective
epochs where $f_{\rm conv}$ is high, the field maintains its polarity and pulsates with a period of $\sim10$ orbits.  In fact, the
only time field reversals occur is when $f_{\rm conv}$ dips to low values\footnote{As discussed in \citet{HIR14} $f_{\rm conv}$ can be slightly negative. This can happen when energy is being advected inwards to the disc midplane.}, indicating
that radiative diffusion is dominating convection.

This lack of field reversals during convective epochs locks the vertical structure of
$B_y$ into either an even parity or odd parity state, where $B_y$ maintains sign across
the midplane or it changes sign, respectively. (Compare orbits 10-40 to orbits 70-100
in the bottom left panel of Fig.~\ref{fig:By}). This phenomenon of the parity of
$B_y$ being held fixed throughout a convective epoch shall henceforth be referred to as
\textit{parity locking}.
During even parity epochs (e.g. orbits 10-40 and 120-140 of ws0446), there are field
reversals in the midplane, but they are quickly quenched, and what field concentrations
are generated here do not migrate away from the midplane as they do during radiative
epochs.
Also during even parity convective epochs, the Poynting flux tends to be oriented inwards roughly half way between the photospheres and the midplane. For odd parity convective epochs the behavior of the Poynting flux is more complicated but is likely linked to the
motion of the $\have{B_y}=0$ surface.

In summary, we seek to explain the following ways in which convection alters the
butterfly diagram:

\begin{enumerate}
\item Magnetic field reversals near the midplane are quickly quenched
during convective epochs.
\item Magnetic field concentrations do not migrate away
from the midplane during convective epochs as they do during radiative epochs.
\item During convective epochs, the magnetic field in the
wings of the butterfly diagram
is better characterized by quasi-periodic pulsations, rather
than quasi-periodic field reversals, with roughly the same period.
\item $B_y$ is held fixed in either an odd or even parity state during convective
epochs.
\end{enumerate}

\section{Thermodynamics and Magnetic Buoyancy}

Much like in \citet{BLA11}, we find that that during radiative epochs, nonlinear
concentrations of magnetic field form in the midplane regions, and these concentrations
are underdense and therefore buoyant.  The resulting upward motion of these field
concentrations is the likely cause of the vertically outward moving field pattern observed in
the standard butterfly diagram.   In our simulations, this magnetic buoyancy appears to
be more important when radiative diffusion, rather than convection, is the predominate
energy transport process.  This is due to the different opacities and rates of radiative
diffusion between these two regimes, which alter the thermodynamic conditions of the
plasma.

When the disc is not overly opaque, and convection is therefore never present, temperature
variations ($\delta T$) at a given height are rapidly suppressed by radiative diffusion.
This causes horizontal variations in gas pressure ($\delta P_{\rm gas}$) and mass density ($\delta \rho$) to be highly correlated (see Fig. \ref{fig:pgas-rho-rad}), and allows
us to simplify our analysis by assuming $\delta T=0$.
This should be contrasted with convective simulations
(see Fig. \ref{fig:pgas-rho-conv}) which show a much noisier relation and show a tendency towards adiabatic fluctuations during convective epochs.

By computing rough estimates of the thermal time we can see how isothermal and adiabatic
behaviour arise for radiative and convective epochs, respectively.
The time scale to smooth out temperature fluctuations over a length scale $\Delta L$
is simply the photon diffusion time times the ratio of gas internal energy density
$e$ to photon energy density $E$,
\begin{equation}
t_{\rm th}\simeq\frac{3\kappa_{\rm R}\rho(\Delta L)^2}{c}\frac{e}{E}.
\end{equation}
For the midplane regions of the radiative simulation ws0429 at times $75-100$ orbits,
the density $\rho\simeq7\times10^{-7}$g~cm$^{-3}$,
$e\simeq2\times 10^7$~erg~cm$^{-3}$, $E\simeq9\times10^5$~erg~cm$^{-3}$, and
the Rosseland mean opacity $\kappa_R\simeq10$~cm$^2$~g$^{-1}$.
Hence, $t_{\rm th}\simeq30(\Delta L/H)^2$~orbits.  Radiative diffusion is therefore
extremely fast in smoothing out temperature fluctuations on scales of order several
tenths of a scale height, and thus horizontal fluctuations are roughly isothermal.

\begin{figure}
	\includegraphics[width=1.0\linewidth]{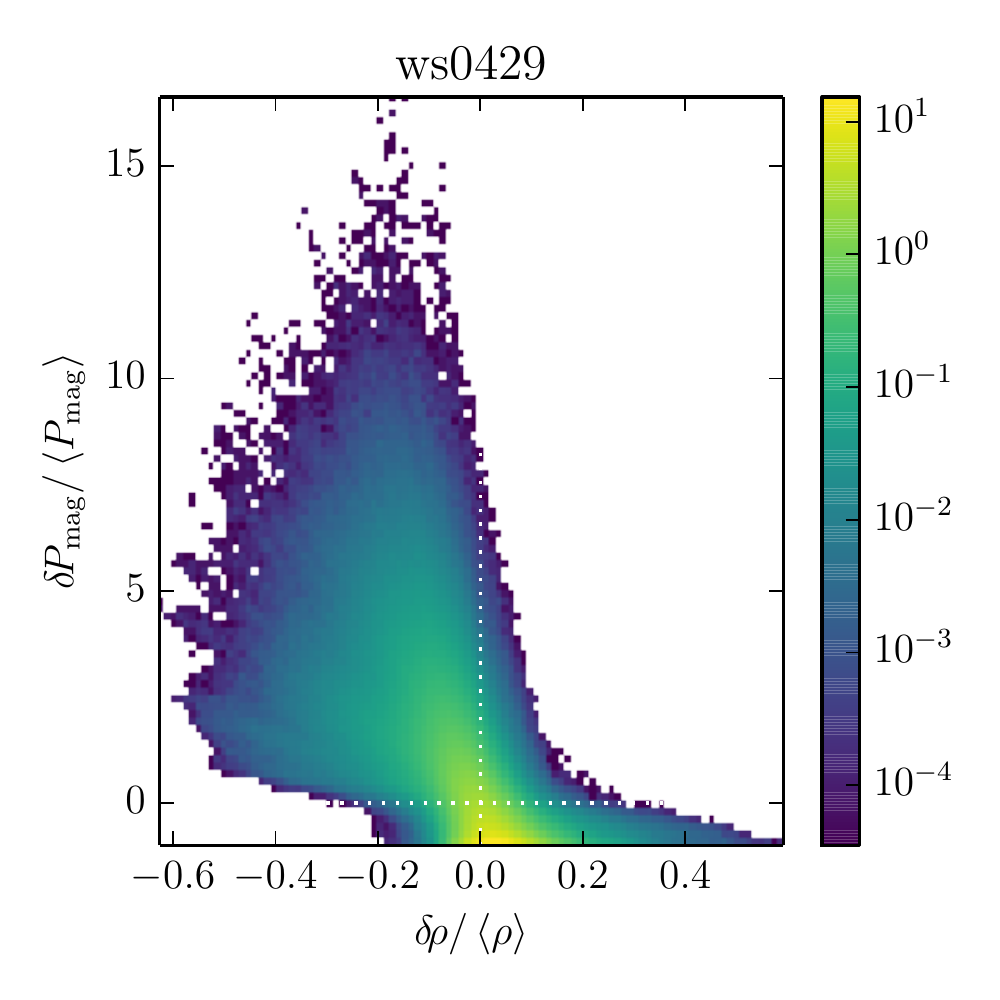}
	\caption{
		A 2D histogram of horizontal variations of mass density ($\delta\rho$, horizontal axis) and magnetic pressure ($\delta P_{\rm mag}$, vertical axis), using data from the same vertical and temporal ranges as Fig.~\ref{fig:pgas-rho-rad}.
		The rough anti-correlation between $\delta\rho$ and $\delta P_{\rm mag}$ signifies that highly magnetized regions are likely to be underdense and thus buoyant.
	}
	\label{fig:pmag-rho-rad}
\end{figure}

Isothermality ($T=\have{T}$)
in combination with pressure equilibrium ($P_{\rm tot}=\have{P_{\rm tot}}$) leads to the following
equation:
\begin{equation}
\label{eqn:isoTP}
\have{P_{\rm tot}}=\dfrac{\rho}{\mu m_p}k\have{T}+P_{\rm mag},
\end{equation}
where radiation pressure has been neglected, as $P_{\rm rad}\ll P_{\rm gas}$.
Thus, during radiative epochs, it is clear that regions of highly concentrated magnetic
field (e.g. flux tubes) must be under-dense. Figure~\ref{fig:pmag-rho-rad} confirms this
for the radiative simulation ws0429 by depicting a 2D histogram of magnetic pressure
and density fluctuations.  A clear anticorrelation is seen which extends up to very
nonlinear concentrations of magnetic field, all of which are underdense.  This
anticorrelation was also observed in radiation pressure dominated simulations
appropriate for high luminosity black hole accretion discs in \citet{BLA11}.
This anti-correlation causes the buoyant rise of magnetic field which would explain
the outward propagation seen in the butterfly diagram and is also consistent with the
vertically outward Poynting flux (see top panels of Figure \ref{fig:By}).

On the other hand, for the midplane regions of the convective simulation ws0446 at
the times $80-100$ orbits, $\rho\simeq2\times10^{-7}$g~cm$^{-3}$,
$e\simeq3\times10^6$~erg~cm$^{-3}$, $E\simeq1\times10^3$~erg~cm$^{-3}$, and
$\kappa_{\rm R}\simeq7\times10^2$~cm$^2$~g$^{-1}$.
Hence, $t_{\rm th}\simeq4\times10^4(\Delta L/H)^2$ orbits.  All fluctuations in
the midplane regions that are resolvable by the simulation are therefore roughly adiabatic.
Perhaps somewhat coincidentally, $\Gamma_1\approx1.3$ in the midplane regions of
the convective simulation, so the pressure-density fluctuations, even though
adiabatic, are in any case close to an isothermal relationship\footnote{
This reduction in the adiabatic gradient within the hydrogen ionization transition
actually contributes significantly to establishing a convectively unstable situation
in our dwarf nova simulations.  We typically find that the adiabatic temperature gradient
$\nabla_{\rm ad}$ within
the hydrogen ionization transition is significantly less than the value 0.4 for a monatomic gas.  In fact, the gas pressure weighted average value of
$\nabla_{\rm ad}$ can be as low as 0.18.  For $\sim 60\%$ of the convective simulations
of \citet{HIR14} and \citet{COL16}, the temperature gradient $\nabla$ is superadiabatic
but less than 0.4 during convective epochs.}.
However, the biggest difference between the radiative and convective cases
is caused by the departure from isothermality in convective epochs, allowing for the possibility of highly magnetized regions to be overdense. This leads to much larger scatter in the probability distribution of the density perturbations in convective epochs.
How this affects magnetic buoyancy in
radiative and
convective epochs will be discussed
in detail in the next section.

\section{Effects of Convection}

In this section we lay out the main mechanisms by which convection acts to modify the
dynamics of the dynamo and thereby fundamentally alter the large scale magnetic field
structure in the simulations.

\subsection{Mixing from the Wings}

\begin{figure}
	\includegraphics[width=1.0\linewidth]{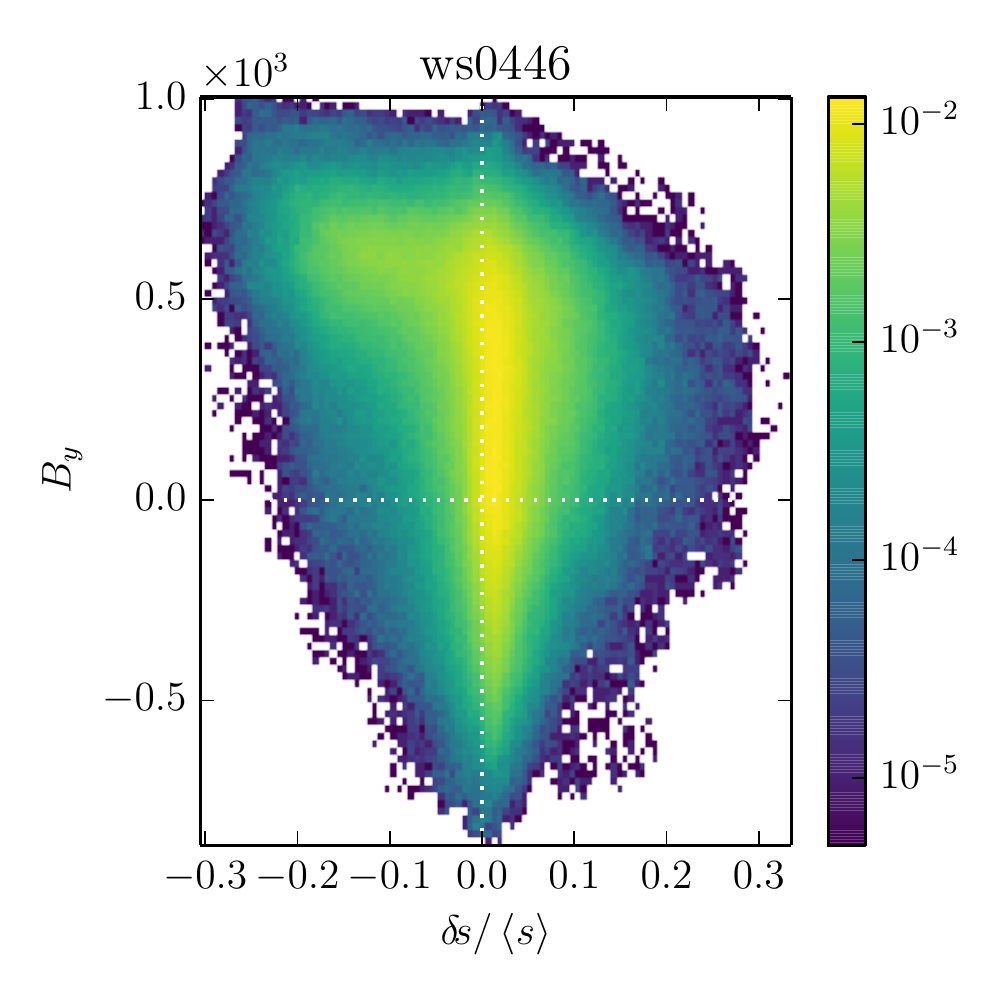}
	\caption{
		A 2D histogram of azimuthal magnetic field ($B_y$, vertical axis) and horizontal variations of specific entropy ($\delta s$, horizontal axis) for an even parity
convective epoch.
		Each cell saved to file from ws0446 with $15\leq t \leq40$ and $|z|\leq 0.25$ simulation units is placed into one of $100\times 100$ bins based on its local properties. The light green region at low entropy ($-0.2\lesssim \delta s < 0$) and high magnetic field ($B_y\gtrsim 500$) signifies highly magnetized fluid parcels which have been mixed towards the midplane by convection.
	}
	\label{fig:even}
\end{figure}

\begin{figure*}
	\includegraphics[width=1.0\linewidth]{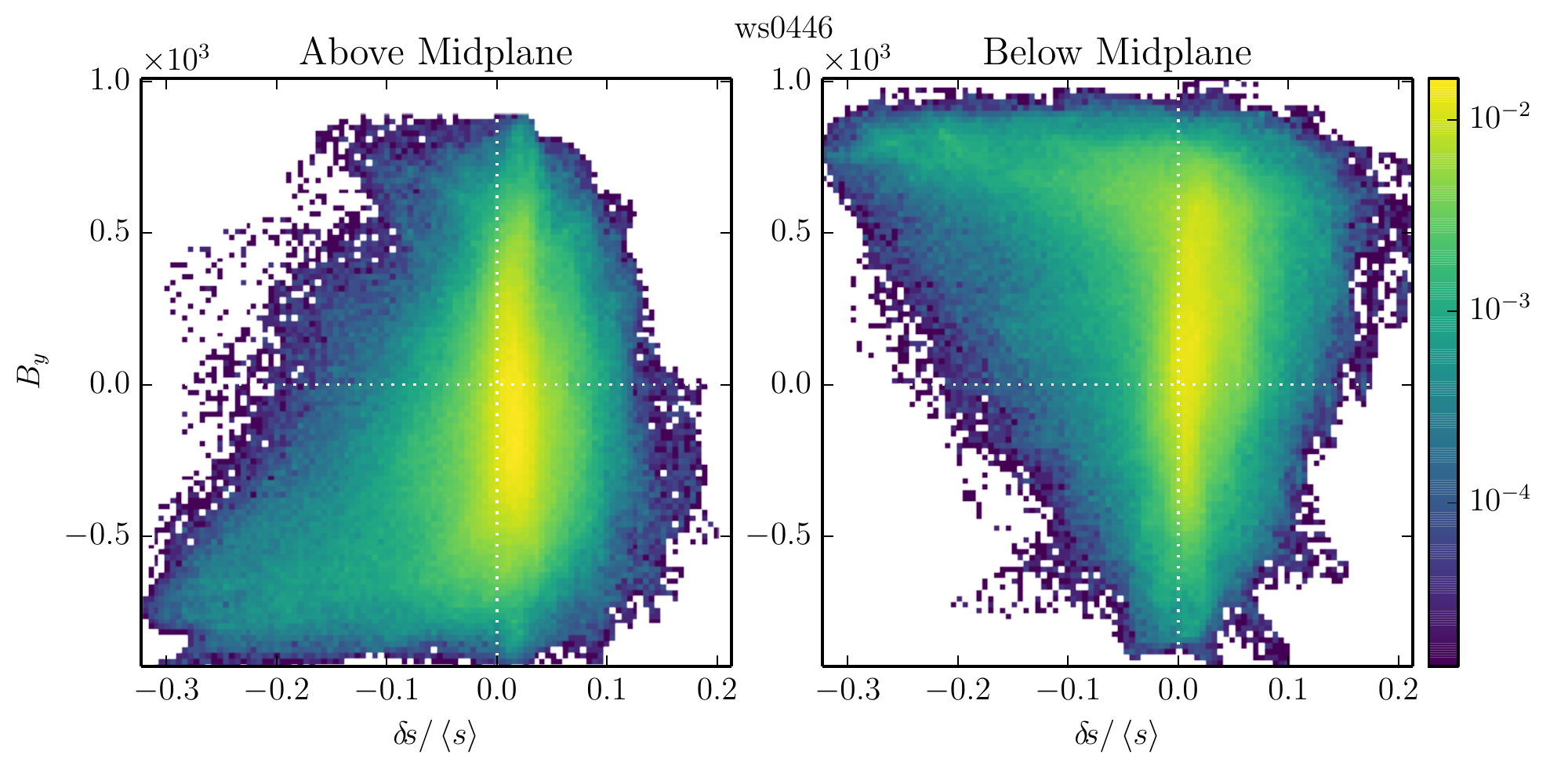}
	\caption{
		2D histograms of azimuthal magnetic field ($B_y$, vertical axis) and horizontal variations of specific entropy ($\delta s$, horizontal axis).
		Each cell saved to file from ws0446 with $80\leq t \leq100$ and $0\leq z\leq 0.25$ simulation units (left) and $-0.25\leq z\leq 0$ simulation units (right) are used to create the respective histograms.
		It is important to note that for both panels there is a low entropy, high $|B_y|$ tail which corresponds to the sign of the respective wings of the butterfly
diagram (see bottom panel of Fig. \ref{fig:By}), indicating that convection is mixing regions of high magnetization from the wings into the midplane.
	}
	\label{fig:odd}
\end{figure*}

As a convective cell brings warm underdense plasma from the midplane outward towards
the photosphere (i.e. the wing region of the butterfly diagram), it must also circulate
cold overdense material from the wing down towards the midplane.
As is typical of stratified shearing box simulations of MRI turbulence (e.g.
\citealt{MIL00,KRO07}), the horizontally and time-averaged magnetic energy density peaks away from the midplane, and the surface photospheric regions are magnetically dominated.
Dense fluid parcels that sink down toward the midplane are therefore likely to carry
significant magnetic field inward.
These fluid parcels that originated from high altitude can actually be identified
in the simulations because the high opacity, which contributes to the onset of
convection, prevents cold fluid parcels from efficiently thermalizing with their
local surroundings.  Hence they retain a lower
specific entropy compared to their surroundings as they are brought to the midplane by
convective motions.  We therefore expect negative specific entropy fluctuations in the
midplane regions to be correlated with high azimuthal magnetic field strength of the
same polarity as the photospheric regions during a convective epoch.
Figures~\ref{fig:even} and \ref{fig:odd} show that this is indeed the case.

Figure~\ref{fig:even} shows a 2D histogram of entropy fluctuations and azimuthal
field strength $B_y$ in the midplane regions for the even parity convective epoch
$15\le t\le40$
in simulation ws0446 (see Fig.~\ref{fig:By}).  The yellow vertical population is
indicative of adiabatic fluctuations (i.e. $\delta s=0$) at every height which are
largely uncorrelated with $B_y$.  However, the upper left quadrant of this figure
shows a significant excess of cells with lower than average entropy for their height
and large positive $B_y$.  It is important to note that this corresponds to the sign of
the azimuthal field at high altitude, even though near the midplane $\have{B_y}$ is
often negative (see bottom panel of Fig. \ref{fig:By}).  This is strong evidence that
convection is advecting low entropy magnetized fluid parcels from the near-photosphere
regions into the midplane.

Figure~\ref{fig:odd} shows the same thing for the odd parity convective epoch
$80\le t\le100$.  Because the overall sign of the horizontally-averaged azimuthal
magnetic field flips across the midplane, cells that are near but above the
midplane are shown in the left panel, while cells that are near but below the midplane
are shown on the right.  Like in Figure~\ref{fig:even}, there is a significant excess
of negative entropy fluctuations that are correlated with azimuthal field strength and
have the same sign as the field at higher altitudes {\it on the same side of the
midplane}.  Again, these low entropy regions represent fluid parcels that have
advected magnetic field inward from higher altitude.  The correlation between
negative entropy fluctuation and azimuthal field strength is somewhat weaker than in
the even parity case (Fig.~\ref{fig:even}), but that is almost certainly due to the
fact that inward moving fluid elements can overshoot across the midplane.

\begin{figure}
	\includegraphics[width=1.0\linewidth]{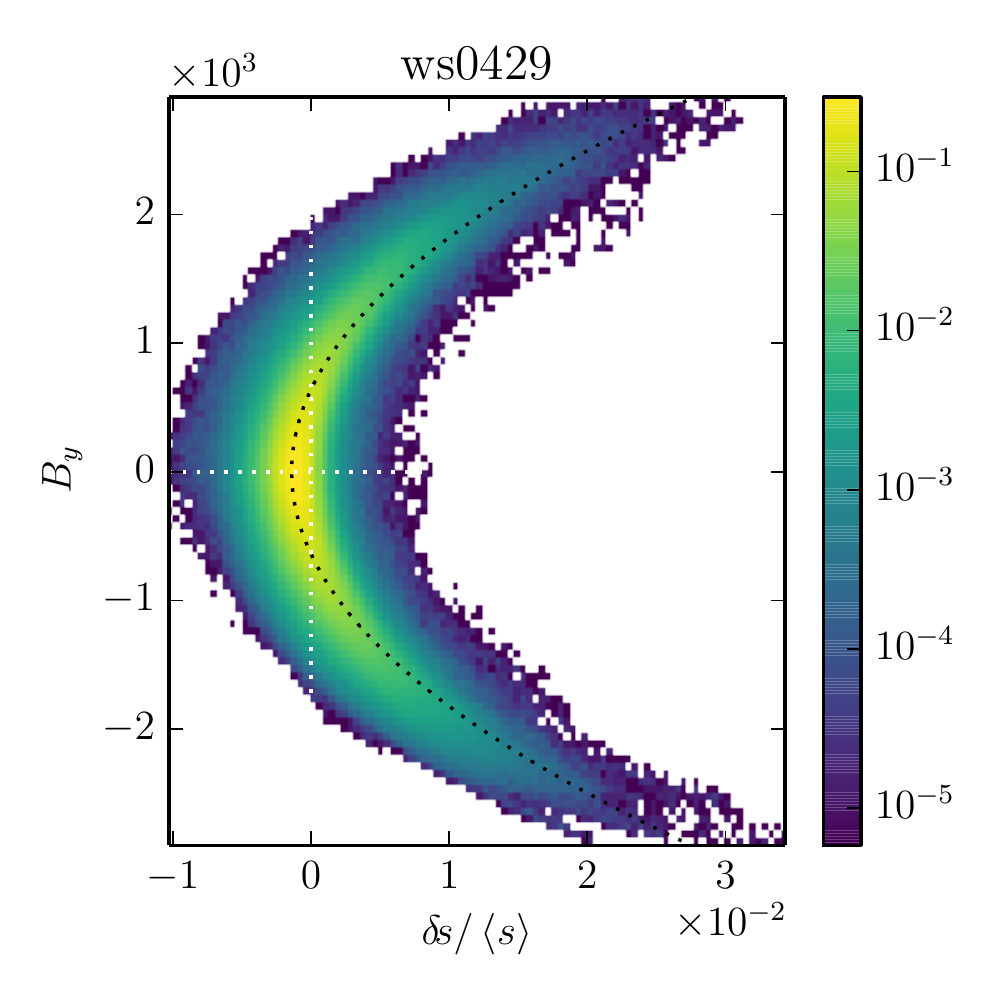}
	\caption{
		A 2D histogram of azimuthal magnetic field ($B_y$, vertical axis) and horizontal variations of specific entropy ($\delta s$, horizontal axis) for the radiative simulation ws0429.
Note the symmetry and how the probability distribution curves to high entropy at high magnetic field strength. This curvature is indicative of isothermal, isobaric fluctuations
and matches well with linear theory (black dotted line) which is computed from
Eqn.~\ref{eqn:ds} assuming that $B^2=B_y^2$. This trend should be contrasted with
what we observe during convective epochs in Figs.~\ref{fig:even} and \ref{fig:odd}.
Note that the entropy fluctuations here are an order of magnitude smaller compared to
those of the convective simulation ws0446.
	}
	\label{fig:sB_rad}
\end{figure}

In contrast, Figure~\ref{fig:sB_rad} shows a similar histogram of entropy fluctuations
and $B_y$ for the radiative simulation ws0429, and it completely lacks this correlation
between high azimuthal field strength and negative entropy fluctuation.  This is in
part due to the fact that fluid parcels are no longer adiabatic, but isothermal.  But
more importantly, it is because there is no mixing from the highly magnetized regions
at high altitude down to the midplane.
Instead, the tight crescent shaped correlation of Figure~\ref{fig:sB_rad} arises simply
by considering the linear theory of isothermal, isobaric fluctuations at a particular
height.  Such fluctuations have perturbations in entropy given by
\begin{equation}
\label{eqn:ds}
\delta s =\dfrac{k}{\mu m_p}\dfrac{B^2-\have{B^2}}{2\have{P_{\rm gas}}}.
\end{equation}
This is shown as the dotted line in Figure~\ref{fig:sB_rad}, and fits the observed
correlation very well.

The inward flux of magnetic energy from high altitude is also energetically large enough to quench field reversals in the midplane regions.  To demonstrate this, we examined the divergence of the Poynting flux and compared it to the time derivative of the magnetic pressure.
During radiative epochs when the magnetic field is growing after a field reversal, typical values for ${\rm d}\have{P_{\rm mag}}/{\rm d}t$ in the midplane are about half of the typical value of $-{\rm d}\have{F_{{\rm Poynt},z}}/{\rm d}z$ near the midplane during convective epochs. This shows that the magnetic energy being transported by the Poynting flux during convective epochs is strong enough to quench the field reversals that would otherwise exist. The sign of the divergence of the Poynting flux during convective epochs is also consistent with magnetic energy being removed from high altitude (positive) and deposited in the midplane (negative).

To conclude, by using specific entropy as a proxy for where a fluid parcel was last
in thermal equilibrium, we have shown that convection advects field inward from high
altitude, which is consistent with the inward Poynting flux seen during even parity
convective epochs (see Fig. \ref{fig:By}). The lack of such clear inward Poynting flux
during the odd parity convective epochs is likely related to the movement of the
$\have{B_y}=0$ surface by convective overshoot across the disc midplane.
However, in both even and odd convective epochs
${\rm d}\have{F_{{\rm Poynt},z}}/{\rm d}z$ is typically a few times ${\rm d}\have{P_{\rm mag}}/{\rm d}t$ and is consistent with enough magnetic energy being deposited in the midplane to quench field reversals that would otherwise take place. This further suggests that
regardless of parity,
this convective mixing from high altitude
to the midplane is quenching field reversals in the midplane by
mixing in field of a consistent polarity.

\subsection{Disruption of Magnetic Buoyancy}

\begin{figure*}
	\includegraphics[width=1.0\linewidth]{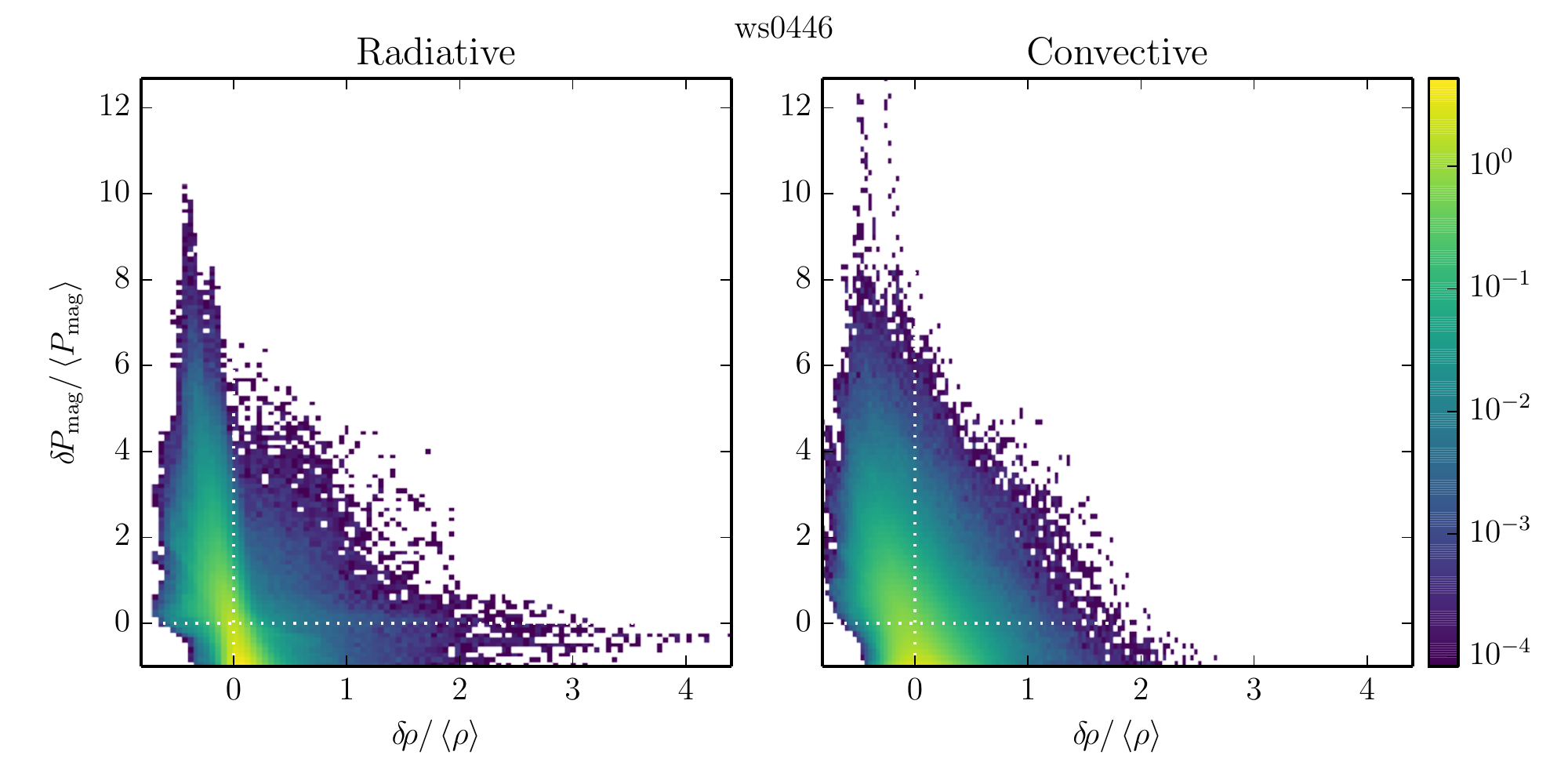}
	\caption{
		2D histograms of horizontal variations of mass density ($\delta\rho$, horizontal axis) and magnetic pressure ($\delta P_{\rm mag}$, vertical axis).
		The left panel shows data from a radiative epoch ($49\leq t \leq 64$) of simulation ws0446 and the right panel
shows data from a convective epoch ($70\leq t \leq 100$) of the same simulation.
		While there is still a rough anti-correlation between $\delta\rho$ and $\delta P_{\rm mag}$ as seen in Fig. \ref{fig:pmag-rho-rad} this correlation is much weaker here. Also, note that the scale of the x-axis here is an order of magnitude higher than that of  Fig. \ref{fig:pmag-rho-rad}. The left frame is similar to the data from ws0429 but has an extra source of dispersion which is likely connected to previous episodes of convection. The right frame has a significantly higher population of overdense and highly magnetized fluid parcels compared to the radiative epoch. This signifies that magnetic buoyancy is weakened during convective epochs.
	}
	\label{fig:pmag-rho-conv}
\end{figure*}

In addition to quenching magnetic field reversals, convection
and the associated high opacities
act to disrupt magnetic buoyancy which transports field away from the midplane, thereby preventing any reversals which do occur, from propagating vertically outwards.
The large opacities which contribute to the onset of convection also allow for thermal fluctuations on a given horizontal slice ($\delta T$) to persist for several orbits. This breaks one of the approximations which lead to the formulation of Eqn. \ref{eqn:isoTP}, allowing for the possibility for large magnetic pressures to be counterbalanced by low temperatures, reducing the anti-correlation between density and temperature.
Additionally, convective turbulence also generates density perturbations that are uncorrelated with magnetic fields, and combined with the lack of isothermality can cause fluid parcels to have both a high magnetic pressure and be overdense (see Fig. \ref{fig:pmag-rho-conv}).
These overdense over-magnetized regions can be seen in the right (convective) frame of Fig. \ref{fig:pmag-rho-conv} where the probability density at
$\delta \rho\approx 0.5\have{\rho},\;\delta P_{\rm mag} \approx\have{P_{\rm mag}}$
is only about one order of magnitude below its peak value. This should be contrasted with the left (radiative) frame of the same figure and with Fig. \ref{fig:pmag-rho-rad} where the probability density at this coordinate is very small or zero respectively.
Hence while the overall anti-correlation between magnetic pressure and density still exists in convective epochs, indicating some magnetic buoyancy, the correlation is weakened by the presence of overdense high magnetic field regions. Magnetic buoyancy is therefore weakened compared to radiative epochs.

\subsection{Parity Locking}

The effects described above both prevent magnetic field reversals in the midplane and reduce the tendency for any reversals which manage to occur from propagating outwards. Therefore convection creates an environment which prevents field reversals, and leads to the parity of the field being locked in place. Due to the variety of parities seen, it appears likely
that it is simply the initial conditions when a convective epoch is initiated that set
the parity for the duration of that epoch.

\section{Discussion}

It is important to note that we still do not understand many aspects of the MRI
turbulent dynamo.  Our analysis here has not shed any light on the actual origin of
field reversals in the standard (non-convective) butterfly diagram, nor have we provided
any explanation for
the quasi-periodicities observed in both the field reversals of the standard diagram
and the pulsations that we observe at high altitude during convective epochs.  However,
the outward moving patterns in the standard butterfly wings combined with the fact
that the horizontally-averaged Poynting flux is directed outward strongly suggests that
field reversals are driven in the midplane first and then propagate out by magnetic
buoyancy.  On the other hand, we continue to see the same quasi-periodicity
at high altitude in convective epochs as we do in the field reversals in the radiative
epochs.  Moreover, it is clear from Figure~\ref{fig:By} that field reversals occasionally
start to manifest in the midplane regions during convective epochs, but they simply
cannot be sustained because they are annihilated by inward advection of magnetic
field of sustained polarity.  This suggests
perhaps that there are two spatially separated dynamos which are operating.
This modification of the dynamo by convection presents a challenge to dynamo models. However, potentially promising dynamo mechanisms have recently been discovered in stratified and unstratified shearing boxes.

Recently, \citet{SHI16} found that quasi-periodic azimuthal field reversals
occur even in unstratified, zero net flux shearing box simulations, provided
they are sufficiently tall ($L_z/L_x \geq 2.5$).
Furthermore, they found
that the magnetic shear-current effect \citep[discussed in][]{S&B_2015} was responsible for
this dynamo; however, they were not able to explain why the reversals occurred.
The shear-current effect can also apparently be present during hydrodynamic
convection \citep{ROJ_2007}, implying that this dynamo mechanism might be
capable of persisting through convective epochs.

Most of the work on the MRI dynamo has been done with vertically stratified
shearing box simulations. One of the earliest examples of this is \citet{BT04}, who found that multiple dynamos can work in conjunction with the MRI on different scales.
This is consistent with the findings of \citet{GRE10}, that an ``indirect'' larger scale dynamo should coexist with the MRI, and they propose two candidates: a Parker-type dynamo \citep[i.e. the $\alpha$-effect;][]{PAR55,RK93}, and a ``buoyant'' dynamo caused
by the Lorentz force. Furthermore, \citet{GRE15} find that the $\alpha\Omega$ dynamo produces cycle frequencies comparable to that of the butterfly diagram, and that there is a non-local relation between electromotive forces and the mean magnetic field which varies vertically throughout the disc. This sort of non-local description may be necessary to understand how the midplane and high altitude regions differ from each other, and we hope to pursue such an analysis in future
work.

\subsection{Departures from Standard Disc Dynamo: Comparison with Other Works}

We find that some properties of the dynamo during convective epochs are similar to the convective simulations of \citet{BOD12,BOD15}, such as prolonged states
of azimuthal magnetic field polarity and an enhancement of Maxwell stresses
compared to purely radiative simulations (see, e.g. Figure 6 of
\citealt{HIR14}).
However, there are some dynamo characteristics observed by \citet{BOD15} that are not present in our simulations.
For example, their simulations typically evolved to  a strongly magnetized
state, something which we never find.
We also
find that in our simulations, which exhibit intermittent convection,
the time-averaged Maxwell stress in the midplane regions is approximately
the same in both the convective and radiative epochs\footnote{Although the
Maxwell stress is approximately the same between radiative and convective epochs
in a given simulation, the $\alpha$ parameter is enhanced during convective
epochs because the medium is cooler and the time-averaged midplane pressure is
smaller.},
and is independent of the vertical parity of the azimuthal field.  In contrast,
\citet{BOD15} find substantially less Maxwell stress during odd parity epochs.
Azimuthal field reversals occasionally occur during their convective simulations, whereas we never see such reversals during our convective epochs.  Finally, their simulations exhibit a strong preference for epochs of even parity, and a lack of quasi-periodic pulsations in the wings of the butterfly diagram.

Remarkably, all of these aforementioned properties of their simulations arise in strongly magnetized shearing box simulations \citep{BAI13, SAL16a}. We suggest here that these properties of the dynamo that \citet{BOD15} attribute to convection are actually a manifestation of strong magnetization. To demonstrate this, we start by noting that the simulations presented in \citet{BOD15} adopted: (1) impenetrable vertical boundary conditions that prevented outflows; thus, trapping magnetic field within the domain, and (2) initial configurations with either zero or non-zero net vertical magnetic flux.

\begin{figure*}
  \begin{center}
  \includegraphics[width=\textwidth]{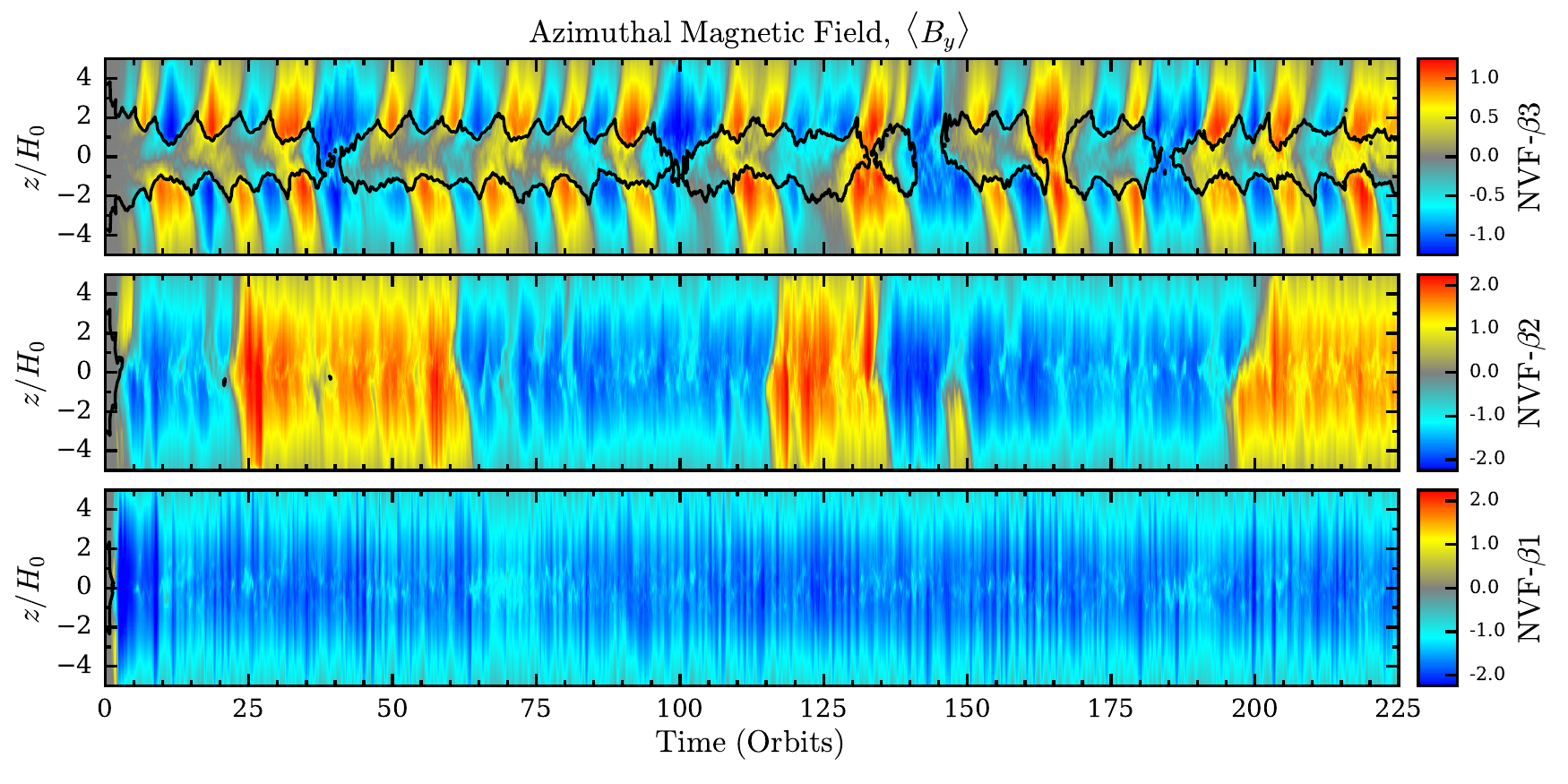}
  \caption{Vertical profiles of the horizontally-averaged azimuthal magnetic field for isothermal shearing box simulations with different levels of net vertical magnetic flux corresponding to an initial mid-plane plasma-$\beta$ of: $\beta_{0}^{\rm mid} = 1000$ ({\it top}), $\beta_{0}^{\rm mid} = 100$ ({\it middle}), $\beta_{0}^{\rm mid} = 10$ ({\it bottom}). {\it Black lines} in the top panel mark the $\beta = 1$ contour and are absent in the bottom two panels because $\beta < 1$ throughout the entire vertical column. For these simulations and the net vertical flux simulations considered by \citet{BOD15} where $\beta_{0}^{\rm mid} \lesssim 1000$, the entire disc becomes strongly magnetized (i.e., $\beta \lesssim 1$ everywhere). The butterfly patterns seen in the convective net vertical flux simulations in Figure 10 of \citet{BOD15} bears a striking resemblance to these isothermal simulations, which do not have convective heat transport. This is suggestive that the dramatic departure from the usual butterfly pattern seen in the net flux simulations of \citet{BOD15} is due to strong magnetization. (This figure is reproduced from \citealt{SAL16a}.)}
  \label{fig:butterfly_netflux}
  \end{center}
\end{figure*}

We first consider the \citet{BOD15} simulations with net vertical magnetic flux. Figure 10 of \citet{BOD15} shows that for increasing net vertical flux, the strength of the azimuthal field increases and field reversals decrease in frequency with long-lived (short-lived/transitionary) epochs of even (odd) parity. No dynamo flips in the azimuthal field were seen for the strongest net flux case. Figure \ref{fig:butterfly_netflux} shows that the {\it isothermal} net vertical flux simulations of \citet{SAL16a} reproduce all features of the butterfly diagrams in the convective simulations of \citet{BOD15}, for the same range of initial plasma-$\beta$. This remarkable similarity between these simulations with and without convective heat transport suggests that strong magnetization (i.e., $\beta \sim 1$ at the disc mid-plane) is responsible for the conflicts listed in the previous paragraph with the \citet{HIR14} simulations under consideration.

We now seek to understand the role of convection on dynamo behavior in the zero net vertical magnetic flux simulations of \citet{BOD15}. These simulations also developed into a strongly magnetized state and exhibited similarly dramatic departures from the standard butterfly pattern as their net flux counterparts. \citet{GRE13} demonstrated that zero net vertical flux shearing box simulations with constant thermal diffusivity and the same impenetrable vertical boundaries adopted by \citet{BOD15} lead to the following: (1) A butterfly pattern that is irregular, yet still similar to the standard pattern that is recovered for outflow boundaries. However, the box size in \citet{GRE13} was comparable to the smallest domain considered in \citet{BOD15}, which was not a converged solution. (2) Maxwell stresses that are enhanced by a factor of $\sim 2$ compared to the simulation with outflow boundaries. This is likely because impenetrable boundaries confine magnetic field, which would otherwise buoyantly escape the domain \citep{SAL16b}. (3) Substantial turbulent convective heat flux, which is significantly reduced when using outflow boundaries. Therefore, perhaps the enhanced convection resulting from using impenetrable boundaries is indeed responsible for the strongly magnetized state and dynamo activity seen in the ``Case D'' zero net flux simulation of \citet{BOD15}.

\begin{figure}
  \begin{center}
  \includegraphics[width=84mm]{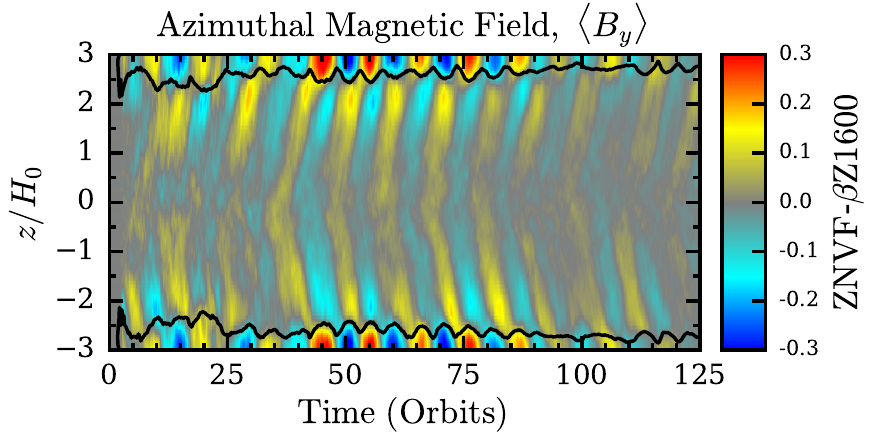}
  \caption{Vertical profile of the horizontally-averaged azimuthal magnetic field for the isothermal, zero net vertical flux shearing box simulation ZNVF-$\beta$1600, which was designed to reproduce the Case D simulation of \citet{BOD15}. Unlike Case D, the ZNVF-$\beta$1600 simulation displays the standard butterfly pattern and does not develop strong magnetization. We note that the \citet{GRE13} simulations with finite thermal diffusivity also did not generate dynamically important magnetic fields.}
  \label{fig:butterfly_Z1600}
  \end{center}
\end{figure}

Despite starting with identical initial and boundary conditions, the zero net vertical flux simulation M4 of \citet{GRE13} does not evolve to the strongly magnetized state seen in the \citet{BOD15} simulations. The reason for this discrepancy is unclear. In an attempt to reproduce Case D in \citet{BOD15} and following \citet{SAL16b},  we ran an {\it isothermal}, zero net vertical flux shearing box simulation that had an initial magnetic field, $\mathbf{B} = B_{0} {\rm sin}\left( 2 \pi x / L_{x} \right)$, where $B_{0}$ corresponded to $\beta_{0} = 1600$. This simulation, labeled ZNVF-$\beta$Z1600, had domain size $\left( L_{x}, L_{y}, L_{z} \right) = \left(24H_{0}, 18H_{0}, 6H_{0}\right)$ with $H_{0}$ being the initial scale height due to thermal pressure support alone, resolution $24~{\rm zones} / H_{0}$ in all dimensions, and periodic vertical boundaries that trap magnetic field, which we believe to be the salient feature of the impenetrable boundaries discussed above. Figure \ref{fig:butterfly_Z1600} shows that the space-time diagram of the horizontally-averaged azimuthal magnetic field for this simulation does not reproduce Case D, but instead evolves to a weakly magnetized state with a conventional butterfly pattern.

However, we note that \citet{GRE13} found that the standard butterfly diagram is recovered when replacing impenetrable boundaries with outflow boundaries. Similarly,
\citet{SAL16b} initialized two zero net vertical flux simulations with a purely azimuthal magnetic field corresponding to $\beta_{0} = 1$. In simulation ZNVF-P, which adopted periodic boundary conditions that prevent magnetic field from buoyantly escaping, the butterfly pattern was not present and the azimuthal field locked into a long-lived, even parity state. However, for simulation ZNVF-O, which adopted outflow vertical boundaries, the initially strong magnetic field buoyantly escaped and the disc settled down to a weakly magnetized configuration with the familiar dynamo activity \citep[see Figure 2 of ][]{SAL16b}. Therefore, vertical boundaries that confine magnetic field may dictate the evolution of shearing box simulations without net poloidal flux.

Based on the discussion above, we suggest that the dynamo behavior in the \citet{BOD15} simulations with net vertical magnetic flux is a consequence of strong magnetization and not convection. For the zero net vertical flux simulations with impenetrable vertical boundaries, the relative roles of convection {\it vs.} strong magnetization in influencing the dynamo is less clear. The main result of this paper --- that convection quenches azimuthal field reversals in accretion disc dynamos --- applies to the case of zero net vertical magnetic flux and realistic outflow vertical boundaries. Future simulations in this regime with larger domain size will help to determine the robustness of this result.

\subsection{Quasi-Periodic Oscillations}	

In addition to the outburst cycles observed in dwarf novae, cataclysmic
variables in general exhibit shorter timescale
variability such as dwarf nova oscillations (DNOs) on $\sim10$~s time scales
and quasi-periodic oscillations (QPOs) on $\sim$ minute to hour
time scales \citep{WAR04}. A plausible explanation for DNOs in CVs involving
the disk/white dwarf boundary layer
has been proposed \citep{WOU02, WAR02}.  However, a substantial number of QPOs
remain unexplained. It is possible that the quasi-periodic
magnetic field reversals seen in the MRI butterfly diagram are responsible for
some of these QPOs and other variability. Temporally, one would expect
variations from the butterfly diagram to occur on minutes to an hour timescales:
\begin{equation}
\tau_{\rm bf} = 223\, {\rm s}\times r_9^{3/2}
\left(\dfrac{M}{0.6\, M_{\sun}}\right)^{-1/2}
\dfrac{\tau_{\rm bf}}{10\,\tau_{\rm orb}},
\end{equation}
where $\tau_{\rm bf}$ is the period of the butterfly cycle, $r_9$ is the radial location of variability in units of $10^9$ cm, $M$ is the mass of the white dwarf primary, and $\tau_{\rm orb}$ is the orbital period.

Indeed, this suggestion has already been made
in the context of black hole X-ray binaries \citep{ONE11,SAL16a}, however, no plausible emission mechanism to convert these field reversals into radiation has been identified. However, it is noteworthy that quasi-periodic azimuthal field
reversals have also been seen in global accretion disk simulations with substantial coherence over
broad ranges of radii \citep[e.g.][]{ONE11,FLO12,JIA14}, indicating that this phenomena is not unique to shearing box simulations.
If QPOs in dwarf novae are in fact associated with azimuthal field reversals,
then our work here further suggests that these QPOs will differ between
quiescence and outburst due to the fact that the convection quenching of field reversals (i.e. the butterfly diagram) only occurs in outburst, and this quenching may leave an observable mark on the variability of dwarf novae. 

\section{Conclusions}

We analyzed the role of convection in altering the dynamo in the shearing box simulations of \citet{HIR14}. Throughout this paper we explained how convection acts to:
\begin{enumerate}
\item quickly quench magnetic field reversals near the midplane;
\item weaken magnetic buoyancy which transports magnetic field concentrations away
from the midplane;
\item prevent quasi-periodic field reversals, leading to quasi-periodic pulsations in the
wings of the butterfly diagram instead; and
\item hold the parity of $B_y$ fixed in either an odd or even state.
\end{enumerate}
All of these are dramatic departures from how the standard quasi-periodic field reversals and resulting butterfly diagram work during radiative epochs.

The primary role of convection in disrupting the butterfly diagram is to mix magnetic field from high altitude (the wings) down into the midplane. This mixing was identified through correlations between entropy and $B_y$ (see Figs. \ref{fig:even}, \ref{fig:odd}). Due to the high opacity of the convective epochs, perturbed fluid parcels maintain their entropy for many dynamical times.  Hence the observed low-entropy highly-magnetized fluid parcels found in the midplane must have been mixed in from the wings. The sign of $B_y$ for these parcels also correspond to the wing which is closest and tend to oppose the sign found in the midplane, quenching field reversals there.

The high opacity which allows for the fluid parcels to preserve their entropy also allows for thermal fluctuations to be long lived. This combined with the turbulence generated by convection weakens the anti-correlation between magnetic pressure and density found in radiative simulations (contrast Figs. \ref{fig:pmag-rho-rad} and \ref{fig:pmag-rho-conv}) and creates an environment where some flux tubes can be overdense. This acts to weaken, but not quench, magnetic buoyancy thereby preventing the weak and infrequent field reversals which do occur from propagating outwards to the wings.

It is through these mechanisms that convection acts to disrupt the butterfly diagram and
prevent field reversals, even though it is clear that the MRI turbulent dynamo continues
to try and drive field reversals in the midplane regions. This results in the sign of
$B_y$ and its parity across the
midplane to be locked in place for the duration of a convective epoch. The quenching of
field reversals and the maintenance of a particular parity (odd or even) across the
midplane is a hallmark of convection in our simulations, and we hope that it may shed
some light on the behaviour of the MRI turbulent dynamo in general.

\section*{Acknowledgements}

We thank the anonymous referee for a constructive report that led to improvements in this paper.
We also wish to acknowledge Tobias Heinemann, Amitiva Bhattacharjee, and Johnathan Squire for their useful discussions and insight generated from their work.
This research was supported by the United States National Science Foundation
under grant AST-1412417 and also in part by PHY11-25915.
We also acknowledge support from the UCSB Academic Senate, and the Center for
Scientific Computing from the CNSI, MRL: an NSF MRSEC (DMR-1121053) and NSF
CNS-0960316.
SH was supported by Japan JSPS KAKENH 15K05040 and the joint research
project of ILE, Osaka University.
GS is supported by an NSF Astronomy and Astrophysics Postdoctoral Fellowship under award AST-1602169.
Numerical calculation was
partly carried out on the Cray XC30 at CfCA, National Astronomical
Observatory of Japan, and  on SR16000 at YITP in Kyoto University.
This work also used the Janus supercomputer, which is supported by the National Science Foundation (award number CNS-0821794) and the University of Colorado Boulder. The Janus supercomputer is a joint effort of the University of Colorado Boulder, the University of Colorado Denver, and the National Center for Atmospheric Research.

\bibliographystyle{apj}
\bibliography{citations}

\begin{thebibliography}{38}
\expandafter\ifx\csname natexlab\endcsname\relax\def\natexlab#1{#1}\fi

\bibitem[{{Bai} \& {Stone}(2013)}]{BAI13}
{Bai}, X.-N., \& {Stone}, J.~M. 2013, \apj, 767, 30

\bibitem[{{Balbus} \& {Hawley}(1998)}]{BAL98}
{Balbus}, S.~A., \& {Hawley}, J.~F. 1998, Reviews of Modern Physics, 70, 1

\bibitem[{{Blackman} \& {Tan}(2004)}]{BT04}
{Blackman}, E.~G., \& {Tan}, J.~C. 2004, \apss, 292, 395

\bibitem[{{Blaes} {et~al.}(2011){Blaes}, {Krolik}, {Hirose}, \&
  {Shabaltas}}]{BLA11}
{Blaes}, O., {Krolik}, J.~H., {Hirose}, S., \& {Shabaltas}, N. 2011, \apj, 733,
  110

\bibitem[{{Bodo} {et~al.}(2012){Bodo}, {Cattaneo}, {Mignone}, \&
  {Rossi}}]{BOD12}
{Bodo}, G., {Cattaneo}, F., {Mignone}, A., \& {Rossi}, P. 2012, \apj, 761, 116

\bibitem[{{Bodo} {et~al.}(2013){Bodo}, {Cattaneo}, {Mignone}, \&
  {Rossi}}]{BOD13}
---. 2013, \apjl, 771, L23

\bibitem[{{Bodo} {et~al.}(2015){Bodo}, {Cattaneo}, {Mignone}, \&
  {Rossi}}]{BOD15}
---. 2015, \apj, 799, 20

\bibitem[{{Brandenburg} {et~al.}(1995){Brandenburg}, {Nordlund}, {Stein}, \&
  {Torkelsson}}]{BRA95}
{Brandenburg}, A., {Nordlund}, A., {Stein}, R.~F., \& {Torkelsson}, U. 1995,
  \apj, 446, 741

\bibitem[{{Brandenburg} \& {Schmitt}(1998)}]{BRA98}
{Brandenburg}, A., \& {Schmitt}, D. 1998, \aap, 338, L55

\bibitem[{{Coleman} {et~al.}(2016){Coleman}, {Kotko}, {Blaes}, {Lasota}, \&
  {Hirose}}]{COL16}
{Coleman}, M.~S.~B., {Kotko}, I., {Blaes}, O., {Lasota}, J.-P., \& {Hirose}, S.
  2016, \mnras, 462, 3710

\bibitem[{{Davis} {et~al.}(2010){Davis}, {Stone}, \& {Pessah}}]{DAV10}
{Davis}, S.~W., {Stone}, J.~M., \& {Pessah}, M.~E. 2010, \apj, 713, 52

\bibitem[{{Flock} {et~al.}(2012){Flock}, {Dzyurkevich}, {Klahr}, {Turner}, \&
  {Henning}}]{FLO12}
{Flock}, M., {Dzyurkevich}, N., {Klahr}, H., {Turner}, N., \& {Henning}, T.
  2012, \apj, 744, 144

\bibitem[{{Galeev} {et~al.}(1979){Galeev}, {Rosner}, \& {Vaiana}}]{GAL79}
{Galeev}, A.~A., {Rosner}, R., \& {Vaiana}, G.~S. 1979, \apj, 229, 318

\bibitem[{{Gressel}(2010)}]{GRE10}
{Gressel}, O. 2010, \mnras, 405, 41

\bibitem[{{Gressel}(2013)}]{GRE13}
---. 2013, \apj, 770, 100

\bibitem[{{Gressel} \& {Pessah}(2015)}]{GRE15}
{Gressel}, O., \& {Pessah}, M.~E. 2015, \apj, 810, 59

\bibitem[{{Hirose}(2015)}]{HIR15}
{Hirose}, S. 2015, \mnras, 448, 3105

\bibitem[{{Hirose} {et~al.}(2014){Hirose}, {Blaes}, {Krolik}, {Coleman}, \&
  {Sano}}]{HIR14}
{Hirose}, S., {Blaes}, O., {Krolik}, J.~H., {Coleman}, M.~S.~B., \& {Sano}, T.
  2014, \apj, 787, 1

\bibitem[{{Hirose} {et~al.}(2006){Hirose}, {Krolik}, \& {Stone}}]{HIR06}
{Hirose}, S., {Krolik}, J.~H., \& {Stone}, J.~M. 2006, \apj, 640, 901

\bibitem[{{Jiang} {et~al.}(2014){Jiang}, {Stone}, \& {Davis}}]{JIA14}
{Jiang}, Y.-F., {Stone}, J.~M., \& {Davis}, S.~W. 2014, \apj, 784, 169

\bibitem[{{Ju} {et~al.}(2016){Ju}, {Stone}, \& {Zhu}}]{JU16}
{Ju}, W., {Stone}, J.~M., \& {Zhu}, Z. 2016, \apj, 823, 81

\bibitem[{{King} {et~al.}(2007){King}, {Pringle}, \& {Livio}}]{KIN07}
{King}, A.~R., {Pringle}, J.~E., \& {Livio}, M. 2007, \mnras, 376, 1740

\bibitem[{{Kotko} \& {Lasota}(2012)}]{K12}
{Kotko}, I., \& {Lasota}, J.-P. 2012, \aap, 545, A115

\bibitem[{{Krolik} {et~al.}(2007){Krolik}, {Hirose}, \& {Blaes}}]{KRO07}
{Krolik}, J.~H., {Hirose}, S., \& {Blaes}, O. 2007, \apj, 664, 1045

\bibitem[{{Miller} \& {Stone}(2000)}]{MIL00}
{Miller}, K.~A., \& {Stone}, J.~M. 2000, \apj, 534, 398

\bibitem[{{O'Neill} {et~al.}(2011){O'Neill}, {Reynolds}, {Miller}, \&
  {Sorathia}}]{ONE11}
{O'Neill}, S.~M., {Reynolds}, C.~S., {Miller}, M.~C., \& {Sorathia}, K.~A.
  2011, \apj, 736, 107

\bibitem[{{Parker}(1955)}]{PAR55}
{Parker}, E.~N. 1955, \apj, 122, 293

\bibitem[{Rogachevskii \& Kleeorin(2007)}]{ROJ_2007}
Rogachevskii, I., \& Kleeorin, N. 2007, Phys. Rev. E, 75, 046305

\bibitem[{{Ruediger} \& {Kichatinov}(1993)}]{RK93}
{Ruediger}, G., \& {Kichatinov}, L.~L. 1993, \aap, 269, 581

\bibitem[{{Salvesen} {et~al.}(2016{\natexlab{a}}){Salvesen}, {Armitage},
  {Simon}, \& {Begelman}}]{SAL16b}
{Salvesen}, G., {Armitage}, P.~J., {Simon}, J.~B., \& {Begelman}, M.~C.
  2016{\natexlab{a}}, \mnras, 460, 3488

\bibitem[{{Salvesen} {et~al.}(2016{\natexlab{b}}){Salvesen}, {Simon},
  {Armitage}, \& {Begelman}}]{SAL16a}
{Salvesen}, G., {Simon}, J.~B., {Armitage}, P.~J., \& {Begelman}, M.~C.
  2016{\natexlab{b}}, \mnras, 457, 857

\bibitem[{{Shakura} \& {Sunyaev}(1973)}]{SS73}
{Shakura}, N.~I., \& {Sunyaev}, R.~A. 1973, \aap, 24, 337

\bibitem[{{Shi} {et~al.}(2010){Shi}, {Krolik}, \& {Hirose}}]{SHI10}
{Shi}, J., {Krolik}, J.~H., \& {Hirose}, S. 2010, \apj, 708, 1716

\bibitem[{{Shi} {et~al.}(2016){Shi}, {Stone}, \& {Huang}}]{SHI16}
{Shi}, J.-M., {Stone}, J.~M., \& {Huang}, C.~X. 2016, \mnras, 456, 2273

\bibitem[{{Squire} \& {Bhattacharjee}(2015)}]{S&B_2015}
{Squire}, J., \& {Bhattacharjee}, A. 2015, Physical Review Letters, 115, 175003

\bibitem[{{Warner}(2004)}]{WAR04}
{Warner}, B. 2004, \pasp, 116, 115

\bibitem[{{Warner} \& {Woudt}(2002)}]{WAR02}
{Warner}, B., \& {Woudt}, P.~A. 2002, \mnras, 335, 84

\bibitem[{{Woudt} \& {Warner}(2002)}]{WOU02}
{Woudt}, P.~A., \& {Warner}, B. 2002, \mnras, 333, 411

\end{thebibliography}

\label{lastpage}

\end{document}